# The VIP Experiment

S. Bartalucci[1], S. Bertolucci[1], M. Bragadireanu[4], C. Bucci[6], M. Cargnelli[3],
M. Catitti[1], C. Curceanu (Petrascu)[1], S. Di Matteo[1], J.-P. Egger[5], N. Ferrari[6],
H. Fuhrmann[3], C. Guaraldo[1], M. Iliescu[1], T. Ishiwatari[3], M. Laubenstein[6],
J. Marton[3], E. Milotti[2,*], D. Pietreanu[1], T. Ponta[4], D. Sirghi[1], F. Sirghi[1],
L. Sperandio[1], E.Widmann[3], J. Zmeskal[3]

presented by Edoardo Milotti

[1] *Laboratori Nazionali di Frascati dell'INFN-LNF, CP 13, Via E. Fermi 40, I-00044 Frascati (Roma), Italy*
[2] *Università Degli Studi di Udine and INFN Sezione di Trieste,*
[3] *"Stefan Meyer" Institute for Subatomic Physics, Boltzmanngasse 3, A-1090 Vienna, Austria*
[4] *"Horia Holubei" National Institute of Physics and Nuclear Engineering, Str. Atomistilor no.407, P.O.BOX MG-6, Bucharest - Magurele, Romania*
[5] *Univ. of Neuchâtel, 1 rue A.-L. Breguet, CH-2000 Fribourg, Switzerland*
[6] *Laboratori Nazionali del Gran Sasso dell'INFN-LNGS, S.S. 17 bis Km. 18.910, 67010 - ASSERGI (AQ), Italy*

**Abstract.** The Pauli Exclusion Principle (PEP) is a basic principle of Quantum Mechanics, and its validity has never been seriously challenged. However, given its importance, it is very important to check it as thoroughly as possible. Here we describe the VIP (Violation of PEP) experiment, an improved version of the Ramberg and Snow experiment (Ramberg and Snow, Phys. Lett. **B238** (1990) 438); VIP shall be performed at the Gran Sasso underground laboratories, and aims to test the Pauli Exclusion Principle for electrons with unprecedented accuracy, down to $\beta^2/2 \approx 10^{-30}$.

**Keywords:** Pauli Exclusion Principle, Symmetrization Principle, Identical Particles

## INTRODUCTION

The Pauli Exclusion Principle (PEP) is a basic principle of Quantum Mechanics, and it is so deeply ingrained in it that it is very difficult even to imagine a formulation of Quantum Mechanics that does not include it. And yet a careful analysis shows that it is not so basic after all, and that in ordinary Quantum Mechanics it is a consequence of a more fundamental principle, the Symmetrization Principle (SP), plus the experimental measurements that fix the symmetry of many-particle wavefunctions. The basic principles of Quantum Mechanics do not force multi-particle states to be either symmetric or antisymmetric, and this requirement must be introduced as an additional principle: using the very words of Messiah, the SP states that "The states of a system containing N identical particles are necessarily either all symmetrical or all antisymmetrical with respect to permutations of the N particles." [1], and it was thoroughly analyzed by Messiah and Greenberg [2]. The situation is different in Quantum Field Theory, where the symmetry of multi-particle states is dictated by the spin-statistics connection, namely the statement that particles with integer spin have symmetrical states and particles with half-integer spin have antisymmetrical states: this was first proved for spin-0 and spin-1/2 particles by Pauli [3], and later on the proof was streamlined by Lüders and

---

[*] Corresponding author: e-mail Edoardo.Milotti@ts.infn.it

Zumino [4] and others (a good starting point to appreciate the problems an difficulties involved is the review [5]), and the spin-statistics connection was extended to arbitrary spins with the introduction of the Schwinger conjecture [6]. Unfortunately the spin-statistics connection lacks an intuitive explanation, and even that great popularizer of the deepest principles of physics, Richard Feynman, gave up and in wrote in his Lectures [7]: "Why is it that particles with half-integral spin are Fermi particles whose amplitudes add with the minus sign, whereas particles with integral spin are Bose particles whose amplitudes add with the positive sign? We apologize for the fact that we cannot give you an elementary explanation. An explanation has been worked out by Pauli from complicated arguments of quantum field theory and relativity. He has shown that the two must necessarily go together, but we have not been able to find a way of reproducing his arguments on an elementary level... . This probably means that we do not have a complete understanding of the fundamental principle involved... .". Given the central standing of these principles, both in Quantum Mechanics and in Quantum Field Theory, the lack of an intuitive explanation, and the possible connection with concepts like the locality of Quantum Field Theory (see later), it is very important to devise tests that may detect small violations for all particle types, and especially for electrons and nucleons [8].

## THE EXPERIMENT OF RAMBERG AND SNOW

There are several kinds of tests that have been reviewed elsewhere [8,9], and here we concentrate on one class of experiments, that was initiated by the work of Goldhaber and Scharff-Goldhaber on the identity of electrons and beta-rays [10]: at that time beta rays were known to be electrons, but all the available experimental evidence could hardly be said to be final and it was open to criticism, so Goldhaber and Scharff-Goldhaber devised an experiment that used PEP to settle this identification problem once and for all. In the experiment, a $C^{14}$ source (in the form of $BaCO_3$ with 3% to 5% $C^{14}$) that emits only beta rays with maximum energy 155 keV and no gamma rays, was used to irradiate a Pb foil: if the beta rays were different from electrons, after being slowed down in the lead foil they would be trapped in an inner level of the Pb atoms, already filled with atomic electrons. Each capture process would be accompanied by the emission of an X-ray, and thus the detection of X-rays by means of a Geiger counter would signal the non-identity of beta-rays and electrons. No emission above the expected background was detected in the experiment which thus established the identity of beta-rays and electrons. Later on, following a suggestion of M. Goldhaber, F. Reines and H. W. Sobel remarked that the same experimental data could be used to set a limit to a conjectured Pauli-violating interaction [11]. Still later, George Snow suggested a modification of the test that used electrons from a current source, rather than beta rays from radioactive decays (the reference to the original suggestion can be found in a paper by Greenberg and Mohapatra [12]). George Snow carried out the experiment with Erik Ramberg (RS), with an apparatus installed on the ground floor of the Muon Building at Fermilab [13] which is schematically shown in figure 1 (see ref. [13] for details).

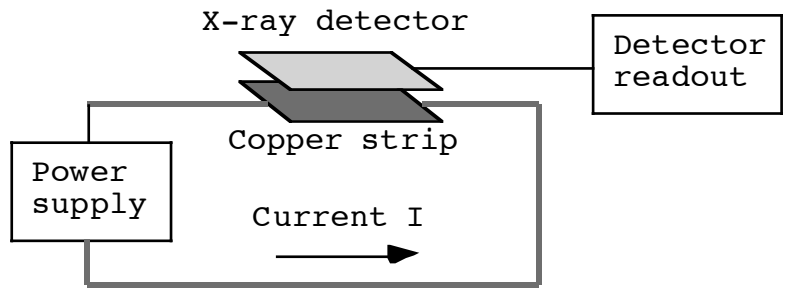

**FIGURE 1.** schematic layout of the Ramberg and Snow experiment: an X-ray sensitive detector is placed close to a copper strip where current from a power supply flows. The measurement is performed both with current on and current off, the resulting X-ray spectra are normalized and the current-off spectrum is subtracted from the current-on spectrum. A Pauli-violating signal would show up as a deviation from the null response in the spectral region close to 7.5 keV (the expected energy of the anomalous X-rays)

According to the suggestion of G. Snow, the "new" electrons are injected by the current source, and the total number of "new" electrons injected into the strip is

$$N_{new} = \frac{1}{e} \int_T I_{vis}(t) dt \qquad (1)$$

where $I_{vis}(t)$ is the fraction of the current that can be observed when one takes X-ray self-absorption in the strip into account and which can be related to the total current $I$, to the the total strip thickness $z$ and to the absorption length $\lambda$ using the formula

$$I_{vis} = I(\lambda/z) = I/(z\sigma\rho) \qquad (2)$$

where $\lambda = 1/\sigma\rho$ ($\sigma$ is the X-ray cross-section for 7.5 keV X-rays in copper, and $\rho$ is the density of copper); the number of electron-atom scattering just in front of the detector is estimated from the formula

$$N_{int} = diameter/m.f.p. = D/\mu \qquad (3)$$

where $D$ is the detector diameter (the detector in the original experiment had a circular window, however $D$ plays the role of a length), and μ is the mean free path for electron-atom scatterings; the normal radiative capture probability is estimated to be about 1/10 of the scattering probability (this can be estimated from the standard radiative width in a hydrogenoid atom and from the estimated dwell time in hopping motion ); and finally

$$N_X \approx \frac{1}{2} \beta^2 N_{new} \frac{N_{int}}{10} \qquad (4)$$

is the expected number of anomalous X-rays (corresponding to the Pauli-violating transition) if $\beta^2/2$ is the Pauli-violating transition probability. Using the values in the RS paper it is possible to set an upper bound for $\beta^2/2$: $\beta^2/2 \leq 1.7 \cdot 10^{-26}$.

## THE VIP EXPERIMENT

The VIP experiment is a new, improved version of the RS experiment, with large-area, sensitive detectors, a low-background experimental area, and a large "electron reservoir". The high efficiency is achieved with scientific-grade CCD detectors fabricated by EEV [14], and already used in the DEAR experiment at the DAFNE machine in the Frascati laboratory of the Italian Institute for Nuclear Physics [15]. A low background is essential to achieve a high sensitivity, because the statistical significance of the subtracted spectrum is a function of the statistical fluctuations, and for this reason the apparatus shall be installed in the Gran Sasso underground laboratory. Additional shielding shall be provided to reject local radioactivity, and we shall flush away radon and any radioactive particulates with a steady flow of dried nitrogen inside the shielding. Figure 2 and 3 show the test setup used to perform preliminary background tests in february 2005. Figure 4 shows the background measured with the test apparatus both in the home Frascati laboratory (which lies in the middle of an ancient volcanic region and is characterized by a high natural radioactive background), and in the Gran Sasso underground laboratory: in the undeground site we have demostrated a 50-fold background reduction with respect to the open-air laboratory, and we hope to achieve a further factor 2 with a more complete shielding, so that in the installed experiment we should have at least a 100 background reduction factor with respect to the open-air laboratory.

The apparatus shall also include an "electron reservoir", i.e. a large block of copper inserted in the current loop that should provide enough "fresh" electrons to let the apparatus run unattended for several months: this is the solid-state equivalent of the beta-ray source in the Goldhaber and Scharff-Goldhaber experiment.

We expect to install the final apparatus during the winter season 2005-6, and immediately start taking data, with a periodic switch between the current-on and current-off status. The experiment should run unattended most of the time, and shall be remotely monitored using a data aquisition system based on the LabView software [16].

Using the new detectors and the large background reduction we hope to achieve a reduction of 3 orders of magnitude on the RS bound on the $\beta^2/2$ parameter.

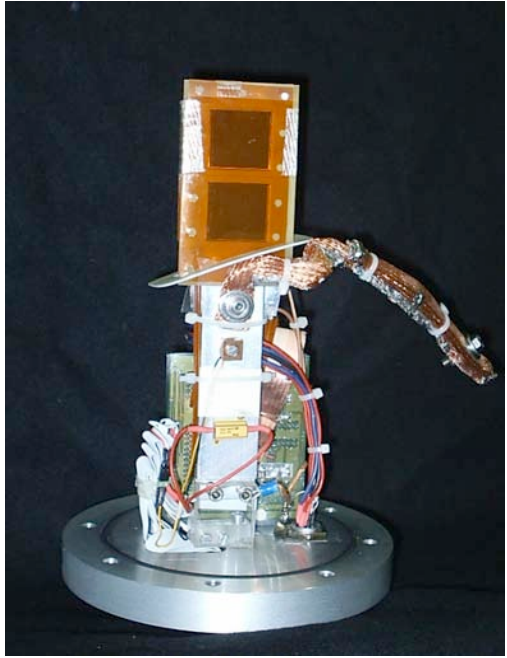

**FIGURE 2**: the 2-CCD test apparatus used to take background data in the Gran Sasso underground laboratory in february 2005. The CCD's are visible in the upper part of the structure, and just below is the control electronics and the support structure that is mantained at low temperature (about 150 K).

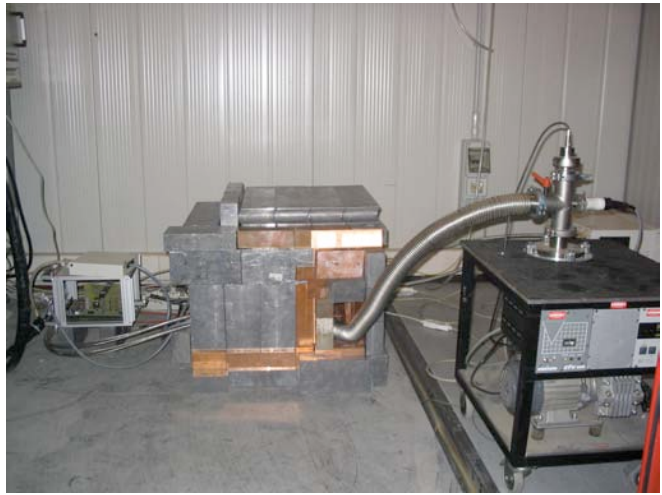

**FIGURE 3**: the test apparatus used to measure the enviromental background in the Gran Sasso laboratory in february 2005. The two-CCD prototype is housed inside the lead- and copper-brick shielding. The CCD's are cooled at about 150 K, and the whole apparatus operates in vacuum: the vacuum pumps are visible on the right.

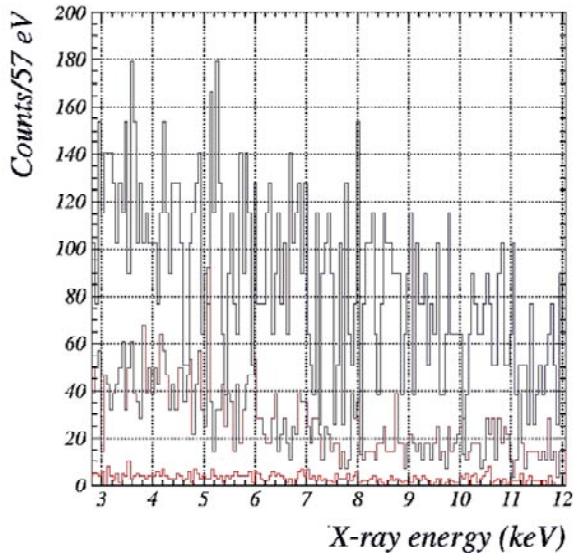

**FIGURE 4**: Measured background counts under different environmental conditions, preliminary tests. Uppermost histogram: in the I.N.F.N. Frascati laboratory, without shielding; middle histogram, same laboratory, with shielding; lowermost histogram: in the Gran Sasso laboratory, with partial shielding.

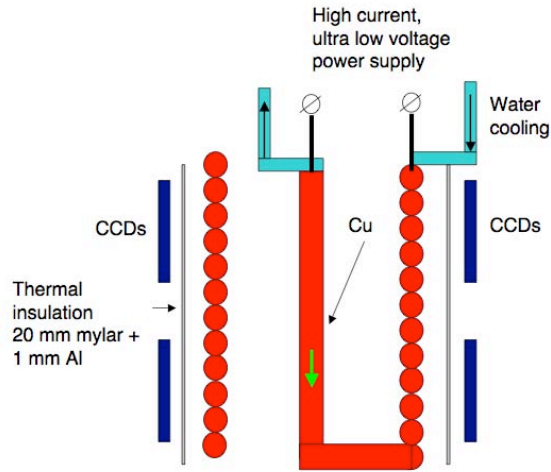

**FIGURE 5**: Proposed layout of the VIP experiment. A water-cooled copper solenoid carries a large current and maximizes the visible current by forcing most of the charge carriers on the surface of the conductor; the forcing is achieved both with proper shaping (the conductor is actually an empty tube) and by the magnetic field (which forces the electrons to stay close to the surface). The X-ray sensitive detectors are positioned around the solenoid on the faces of an octagonal prism, and are thermally isolated with a thin mylar foil (the CCD's must be operated at low temperature, about 150 K, to minimize electrical noise).

## DISCUSSION

As we have seen in the introduction, it is possible to incorporate a violation of the Symmetrization Principle in the usual formalism of Quantum Mechanics: this is not without consequences, a set of single-particle measurements is no longer maximal, and must be supplemented by correlation measurements for a full characterization of the quantum state of the system [2,17]; however this is still tolerable, as long as it does not conflict with the obvious experimental observation that any violation must necessarily be small [12]. Unfortunately it is not straightforward to model the violations dynamically: here we limit our discussion to a few basic facts, and refer the reader to the review papers [2,5,9,12,18] for fuller accounts. In 1987 Ignatiev and Kuzmin cooked up a simple model with three occupation levels [19], and with a modified algebra of the creation and annihilation operators, and have introduced the violation parameter $\beta$ that is used also in the analysis of the Ramberg and Snow experiment. Soon afterwards Govorkov published a series of papers that demostrated that this model is untenable in the framework of standard Quantum Field Theory, unless one gives up some fundamental hypothesis on the structure of spacetime, like locality of creation and annihilation operators [20]. Lev Okun, who published a review on this topic in 1989 [9], now believes that the theory cannot accomodate any violation of the Symmetrization Principle [21]. Greenberg and Mohapatra acknowledged that it is not possible to violate the SP in standard Quantum Field Theory, but believe that accurate tests of the principle are very important [8]. In 1990 and following years, Greenberg and collaborators have constructed a model of particles with weighted commutators

$$\frac{1+q}{2}\left[a_k, a_l^+\right]_- + \frac{1-q}{2}\left[a_k, a_l^+\right]_+ = \delta_{kl} \tag{5}$$

i.e.,

$$a_k a_l^+ - q a_l^+ a_k = \delta_{kl} \tag{6}$$

and have studied in depth this "quon" algebra [22,23]. Quons are well-behaved in many respects [18], but they suffer from problems with many-particle statistics (Gibbs paradox, see ref. [18]), and with locality of creation-annihilation operators [18]. Locality of creation-annihilation operators is one of the basic ingredients of the proof of the spin-statistics connection [4], and violations of locality (and causality) may be considered just bizarre variants of standard phenomenology, but they have a rightful place both in (quasi-)conventional models like the composite electron model of Akama et al. [24], and in those theories that incorporate a holographic principle (if the boundary of space-time determines the structure of the interior of space-time, then events separated by spacelike intervals are actually correlated).

The considerations given above show that, though unlikely, a violation of the SP could be associated to a violation of locality; however, in spite of the great interest that this may have from a theoretical standpoint, it is only vaguely related to an experimental test like VIP, which must rely on a simpler phenomenology. The proper phenomenology is indicated in the papers [2,17,25], and finally it boils down to the analysis performed in the paper of RS.

There are many difficulties in the way of the interpretation of the results: for instance it is not clear which transition should be taken into account to calculate precisely the energy of the anomalous X-rays, i.e., it is unclear at present if the "anomalous" electrons in the conduction band can fall to any occupied level, or if they can only move to a level with a corresponding "paired" anomalous electron.

We conclude with the remark that VIP is a high-precision experiment with a difficult interpretation, that may however shed a little light on the foundations of Quantum Field Theory, and maybe even on the nature of the identity of particles, which could be the key to some of the strangest properties of the quantum world: this is indeed the opinion of the philosopher P. Pesic [26] "Physicists have long struggled with the weirdness of quantum mechanics - a consequence of like particles being completely indistinguishable from one another ...".